**Acquiring Human-Like Data-Efficient Mechanics Prediction from Deep Reinforcement Learning**

Jingruo Peng[1], Shuze Zhu*[1]

[1]Center for X-Mechanics, Institute of Applied Mechanics, Zhejiang University, Hangzhou 310000, China

*To whom correspondence should be addressed. E-mail: shuzezhu@zju.edu.cn

## Abstract

Humans can infer mechanical outcomes by learning from a few observations. This capacity for mechanics intuition is acquired in a data-efficient manner. Here, we propose a reinforcement learning framework to mimic this process, in which an agent encodes continuous physical observation parameters into its state and is trained via episodic switching across closely related observations. With merely two or three similar observations, the agent acquires robust mechanics intuition that generalizes over wide parameter ranges beyond the training data. Our method is demonstrated on the brachistochrone, a large-deformation elastic plate, and the quantum harmonic oscillator. We explain this generalization through a unified theoretical view: it is associated with cross-parameter Bellman consistency encouraged by episodic switching across neighboring task parameters, promoting approximate stationarity of the Bellman residual with respect to physical variations. This is consistent with a smooth policy that tracks a low-dimensional solution manifold underlying the continuum of tasks. Our work identifies episodic switching as a practically effective and theoretically motivated route to artificial mechanics intuition and suggests a computational analogy to data-efficient generalization in biological learners.

## 1. Introduction

Biological organisms acquire intelligent behavior through continual interaction with their environment. Reinforcement learning (RL) algorithms emulate such biological learning by optimizing actions through environmental interaction [1–3]. In humans, this ability to project the consequences of actions without direct trial-and-error is often described as "intuitive reasoning" [4], which in mechanical contexts we refer to as mechanical reasoning or mechanics intuition. For example, when a human mechanics expert observes a mechanical situation of a cantilever beam under loading as described in Figure 1(a), the intuitive understanding on its associated deformation process can be readily formulated as illustrated in Figure 1(b).

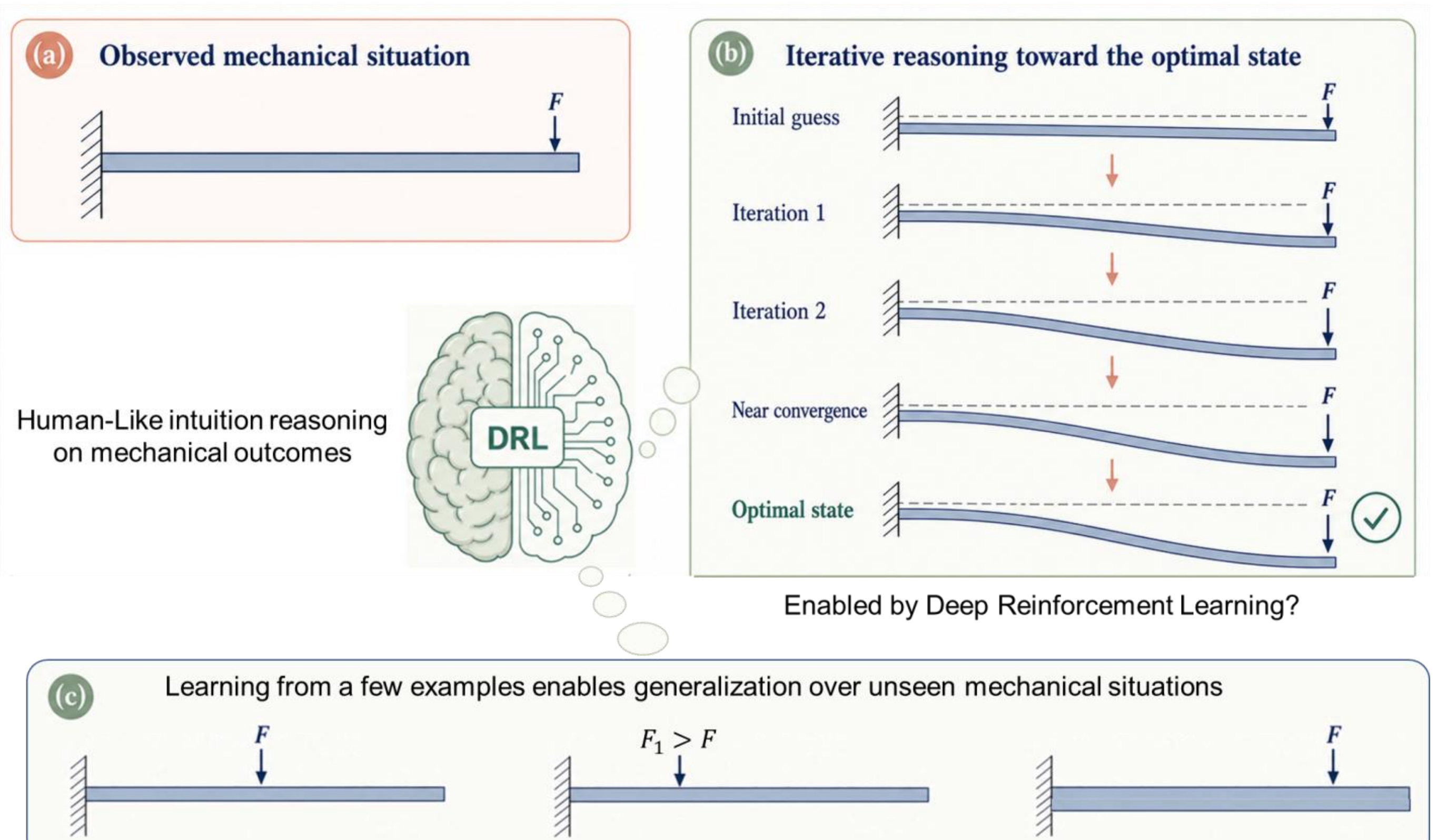

Figure 1. Schematic of Human-like intuition reasoning on mechanical outcomes. (a) Illustration of an observed mechanical situation where a cantilever beam is subjected to a point load. (b) Human expert can reason the evolving step of this mechanical process towards the optimal state. (c) Learning from a few examples, human expert can generalize the knowledge to unseen mechanical situations, making intuitive predictions.

Notably, humans can form robust mechanical intuition from minimal exposure [5–9]. For example, a mason estimates beam deflection by inspection, and a carpenter judges a wooden beam's load-bearing capacity from its grain and geometry. As shown in Figure 1(c), once a human expert gains fine understanding on a single cantilever beam problem, its generalization towards unseen mechanical situations (e.g., different loading location, loading magnitude, beam dimension) is feasible. Cultivating such human-like, data-efficient mechanical insight is important on the path toward artificial general intelligence [9–11].

Currently, the question of whether reinforcement learning is the right approach to artificial intelligence is debated. Critics point to limitations of current RL methods, arguing that richer priors, embodiment, or hybrid architectures are needed for general intelligence [12–16]. Yet, RL has also shown remarkable data-efficiency in domains like language modeling, where human feedback enables smaller models to outperform much larger ones [17–21]. Motivated by these potentials [3], we ask: can an RL agent develop human-like mechanics intuition from extremely limited observations?

To address the above challenge while mimicking the human-like intuition reasoning on mechanical outcomes, we propose a simple yet effective RL framework centered on two core designs: (1) explicit encoding of task-defining physical parameters (hereafter called physical observation parameters), such as geometry and load, into the agent's state representation, structuring the input to reflect the underlying parameter manifold; and (2) episodic observation-switching, a training protocol that repeatedly alternates the agent between tasks with neighboring parameters, encouraging a single strategy that generalizes across local regions of this manifold. We adopt a train–freeze–execute scheme to rigorously evaluate generalization beyond the few observed task instances.

We perform computational experiments on the brachistochrone problem, a large-deformation elastic plate, and the quantum harmonic oscillator, as they are more complex than the simple cantilever illustration in Figure 1. We show that switching across merely two or three similar observations yields a substantial expansion of the high-accuracy region compared to single-observation training. Theoretically, we explain this gain through Bellman consistency: episodic switching encourages smooth variation of the Bellman residual across parameters, which acts as an implicit regularizer, coupling policies and enabling robust extrapolation.

This study demonstrates that RL, when suitably structured, can acquire a form of mechanics intuition from scarce data. By encoding parameters and switching between related tasks, our method exploits inter-task similarity to induce data-efficient generalization. Practically, it offers a pathway to reduce data needs in physics-based RL; conceptually, it bridges data-efficient biological learning with algorithmic designs, showing how architectural and curricular choices can induce useful inductive biases.

## 2. Formulating Mechanics Problems as Reinforcement Learning Tasks

This section provides a complete, self-contained pipeline for converting mechanics problems into RL tasks and obtaining generalizable solutions through structured representation and episodic switching. The final solution is produced by direct rollout of the learned policy, without invoking a task-specific external optimizer at test time, offering a computationally efficient route to data-driven physical intuition.

### 2.1 Problem Formulation as a Markov Decision Process

We formulate variational physical optimization problems (such as finding the brachistochrone curve or the equilibrium displacement field of an elastic plate, or an energy-minimizing wavefunction) as fully observable, deterministic Markov Decision Processes (MDPs). The agent interacts with a physics simulator that embodies the governing mechanics functionals. The objective is to minimize a scalar physical quantity $J$ (e.g., total travel time $T$, the total potential energy $E$, or the Hamiltonian expectation value) through a sequence of actions that incrementally modify the candidate solution.

### 2.2 State Representation: Encoding Physical Observation Parameters

To condition the agent on the specific problem instance, we design a structured state representation that explicitly encodes continuous physical observation parameters $\zeta$. At each collocation point $i$, the state vector is:

$$s_i = [x_i, y_i, \dots, \zeta_1, \zeta_2, \dots]^\top$$

where $(x_i,\ y_i)$ are generalized coordinates, and the parameters $\zeta$ (e.g., endpoint coordinates $\tilde{X}$, $\tilde{H}$, load magnitude $\tilde{P}$, plate size $\tilde{B}$, orientation $\tilde{\Theta}$, or oscillator mass $\tilde{m}$ and angular frequency $\tilde{\omega}$) are repeated across all spatial locations as separate channels (Figure 2(a)). In the following, "physical observation parameter" and "task parameter" are used interchangeably to denote $\zeta$. This structure allows convolutional layers to learn parameter-conditioned spatial patterns while retaining local spatial weight sharing. The set of all $\zeta$ defines a smooth parameter manifold $\mathcal{M}$, where each point corresponds to a task instance. We denote $\zeta$ as the physical observation parameter and use the compact notation $s = (u, \zeta)$ for the state under observation $\zeta$, where $u$ collects the non-parameter state channels, including the current candidate solution and fixed spatial descriptors.

### 2.3 Action Space: Incremental Solution Updates

The action $a_t$ at step $t$ represents one of a finite set of bounded local adjustments to the current solution field. For example, for the brachistochrone, the action modifies the height $y(x)$ at a set of collocation points; for the elastic plate, it updates the displacement components $(U, V, W)$ at each node; for the quantum harmonic oscillator, the action similarly performs bounded local updates to the discretized trial wavefunction. Actions are bounded to ensure physical plausibility and numerical stability. The state transition is deterministic: applying $a_t$ updates the solution field accordingly, and the physics simulator re-evaluates the objective $J$.

### 2.4 Reward Design: Embedding the Physical Objective

The implemented reward is a bounded improvement signal derived from the physical objective. It aligns the training feedback with objective reduction, whereas the smooth perturbation analysis in Section 4 is formulated for the underlying continuous improvement functional, or for a smoothed bounded surrogate of it. The value of the physical objective at the previous and current states is used as feedback from the environment. For the brachistochrone, the reward is defined as $r_t = \text{sign}(T_{t-1} - T_t)$, thereby encouraging reduction of travel time. For the elastic plate, the reward is defined as $r_t = \text{sign}(E_{t-1} - E_t)$, thereby promoting reduction of the total potential energy. For the quantum harmonic oscillator, the reward is defined as $r_t = \text{sign}(\langle E \rangle_{t-1} - \langle E \rangle_t)$, thereby encouraging reduction of the Hamiltonian expectation value and convergence toward the variational ground-state solution. Thus, each improvement in the physical objective yields a positive reward, aligning the RL agent's goal with the principle of least time or the variational equilibrium principle.

### 2.5 Episodic Observation-Switching Protocol

To encourage generalization across nearby tasks, we employ an episodic observation-switching scheme (Figure 2(b)). A key feature is that the network parameters are continuously inherited during episodic observation-switching. Each training episode is associated with one fixed observation parameter value $\zeta_i$. Upon episode termination, the environment switches to a different observation parameter $\zeta_j$ from the same local cluster (with $|\zeta_i - \zeta_j|$ small). The reward function is evaluated using the active observation parameter $\zeta$ of the current episode. This encourages the agent to learn policies that are effective across a neighborhood of parameters rather than overfitting to a single instance.

### 2.6 Network Architecture and Training

We use a dueling Deep Q-Network (DQN) with a convolutional backbone. The network takes the multi-channel state $s$ and outputs the action-value function $Q_\theta(s,a)$. For the same task, all models share the same architecture and hyperparameters. Training is performed on a small, fixed set of observation parameter values $\zeta_1, \zeta_2, \zeta_3$ using experience replay and the switching protocol described above. Specifically, we operate on six training sets: three single-observation models (each trained on one observation $\zeta_i$, labeled as Single $\zeta_i$), two double-observation models (each trained on a pair $\zeta_i, \zeta_j$ via the switching protocol, labeled as Double $\zeta_i, \zeta_j$), and one triple-observation model (trained on all three observations, labeled Triple $\zeta_1, \zeta_2, \zeta_3$). The proximity (i.e., small $|\zeta_i - \zeta_j|$) of these selected parameters within the local cluster is fundamental. It embodies our core assumption: within a sufficiently small neighborhood of the parameter manifold, the optimal value function $Q^*(u, \zeta, a)$ and the corresponding policy vary smoothly. Our episodic switching protocol is explicitly designed to exploit and encourage this local smoothness prior, incorporating it into the training process as an implicit inductive bias.

### 2.7 Obtaining the Final Solution: Direct Execution of the Learned Policy

After training, the network parameters $\theta$ are frozen. The solution to a new problem instance with parameter $\zeta_{test}$ is obtained by directly executing the learned policy. Starting from an initial solution guess (e.g., a straight line or a flat plate), the frozen agent is executed with $\varepsilon = 0$ and follows the greedy policy derived from $Q_\theta$ : $a_t = \arg\max_a Q_\theta(u_t, \zeta_{test}, a)$. For each independent test point, only one deterministic rollout is performed over the fixed interaction budget. The state with the lowest physical objective value $J$ along this single rollout (e.g., travel time for the brachistochrone) is selected as the predicted solution. This selection is an evaluation-time extraction step aligned with the underlying physical objective, whereas the Bellman objective governs the discounted sequential improvement problem used to train the policy. No additional task-specific optimization or fine-tuning is applied. This "train–freeze–execute" pipeline ensures that generalization is evaluated purely on the basis of the learned policy, mimicking how a human expert would intuitively infer a solution across a broad range of previously unseen conditions from prior analogous experience (Figure 2(c)).

### 2.8 Evaluation of Generalization

Generalization is quantified by applying the frozen policy to a dense grid of unseen parameters $\zeta_{test}$ spanning a wide range. For each test case, the predicted solution is compared with a high-fidelity reference (analytic solution or finite-element solution) using the coefficient of determination $R^2$, or the mean $\bar{R}^2$ over multiple displacement components when applicable. A test point is considered successfully generalized if $R^2 \geq 0.90$ or $\bar{R}^2 \geq 0.90$. The union of all such points defines the generalized region in parameter space; its area or interval length, relative to the sparse number of training observations, characterizes the method's data efficiency and generalization capacity.

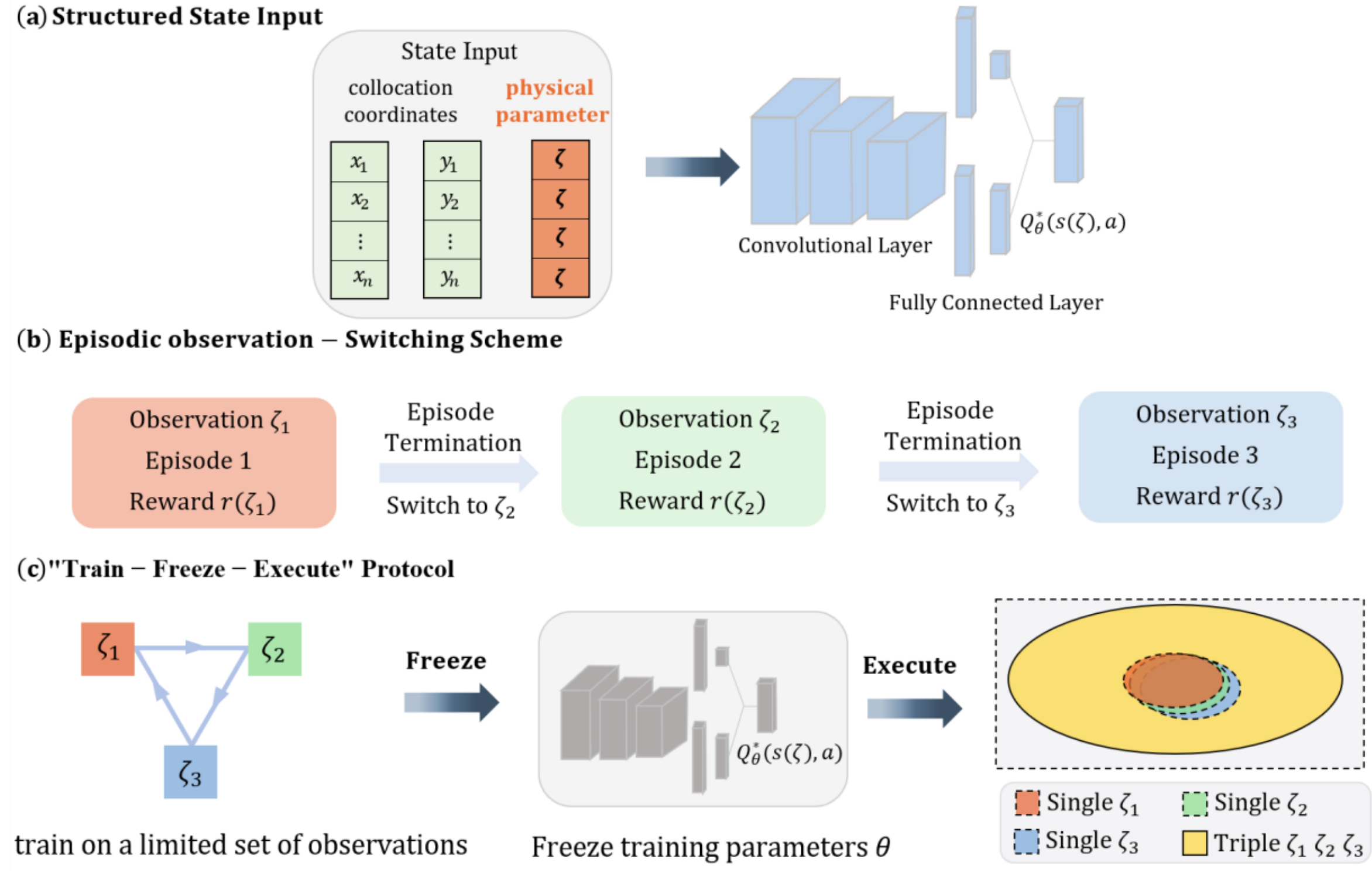

Figure 2. Schematic of episodic observation-switching training with structured state inputs. (a) Structured state input for continuous-field problems. The collocation generalized coordinates $(x_i, y_i)$ and physical observation parameters $\zeta$ are encoded into dedicated numerical channels and fed into a dueling Q-network to output the action–value estimates $Q_\theta(s, a)$, equivalently $Q_\theta(u, \zeta, a)$. (b) Episodic observation-switching training scheme. Each episode is associated with a fixed observation $\zeta_i$, and the observation is switched to $\zeta_j$ upon episode termination; the reward at each episode is defined as a function of the physical observation parameter associated with the active observation. (c) Train–freeze–execute protocol: the network is trained on a set of observations (e.g., $\zeta_1, \zeta_2, \zeta_3$), after which the learned parameters $\theta$ are frozen and the generalization region is evaluated over an extended parameter range, illustrating the evaluation of enlarged high-accuracy regions compared with single-observation training.

## 3. Computational Experiments

### 3.1 Brachistochrone

We validate our "train–freeze–execute" protocol on a canonical mechanics benchmark: the brachistochrone problem. In this task, the agent predicts the time-optimal descent curve for a point particle subject to gravity, traveling from the fixed start $(0,0)$ to a parametrized endpoint $(\tilde{X}, \tilde{H})$. The analytic time-optimal solution is the cycloid, which serves as our reference (Figure 3(a)). The brachistochrone problem is formulated in nondimensional units, and $\tilde{X}$ and $\tilde{H}$ denote nondimensional endpoint coordinates. After each action during the training rollout, we compute the total traversal time $T$ for the current candidate curve by numerical quadrature so that $T = \int_{x_0}^{\tilde{X}} \sqrt{\frac{1+(y')^2}{2gy(x)}}\, dx$ (with $y(x)$ measured as vertical drop). In our RL framework the physical objective functional is encoded into the reward. The agent proposes an incremental modification of the path and the total travel time $T_t$ is recalculated. The reward is defined as $r_t = \mathrm{sign}(T_{t-1} - T_t)$, ensuring that updates that reduce time are positively reinforced. To allow the network to condition on the endpoint geometry, we encode the target endpoint coordinates $(\tilde{X}, \tilde{H})$ as two separate state channels. Each collocation point is therefore represented by a four-channel state vector, $s = (x, y, \tilde{X}, \tilde{H})$, where $(x, y)$ are the local spatial coordinates of the collocation points and $(\tilde{X}, \tilde{H})$ are constant across the nodes but presented as explicit input channels so the agent can learn endpoint-conditioned policies (Figure 3(b)).

We form local clusters of three neighboring observations corresponding to the physical observation parameter triplet $\zeta_1, \zeta_2, \zeta_3$. We compare six variants that share the same DQN backbone and hyperparameter set. After training, we freeze the network parameters and evaluate performance over a dense grid of unseen terminal points $(\tilde{X}, \tilde{H})$ spanning the two-dimensional parameter space. For every test endpoint we compute the network's predicted minimum-time trajectory $y_{pred}$ and compare it to the cycloid $y_{ref}$. A test point is declared "generalized" when the coefficient of determination satisfies $R^2(y_{pred}, y_{ref}) \geq 0.90$. The generalized test points are aggregated to form a generalized region in the $(\tilde{X}, \tilde{H})$ parameter space. Figure 3(c–e) summarizes the generalized regions obtained for all models. The results show that the high-accuracy coverage of the single-observation models is tightly concentrated near the training observations and decreases rapidly with increasing distance. In contrast, models trained on two neighboring observations expand the high-accuracy coverage relative to the union of the corresponding single-observation regions. Incorporating a third neighboring observation further enlarges the generalized region. The triple-observation model attains high-fidelity predictions over a large portion of the evaluation grid, with predictions closely matching the analytic solutions near training observations and predictable accuracy decay toward domain boundaries. Overall, adding neighboring observations enables the agent to yield robust mechanics intuition over a substantially enlarged parameter domain, indicating an expansion of the generalized region beyond the union of the corresponding single-observation regions.

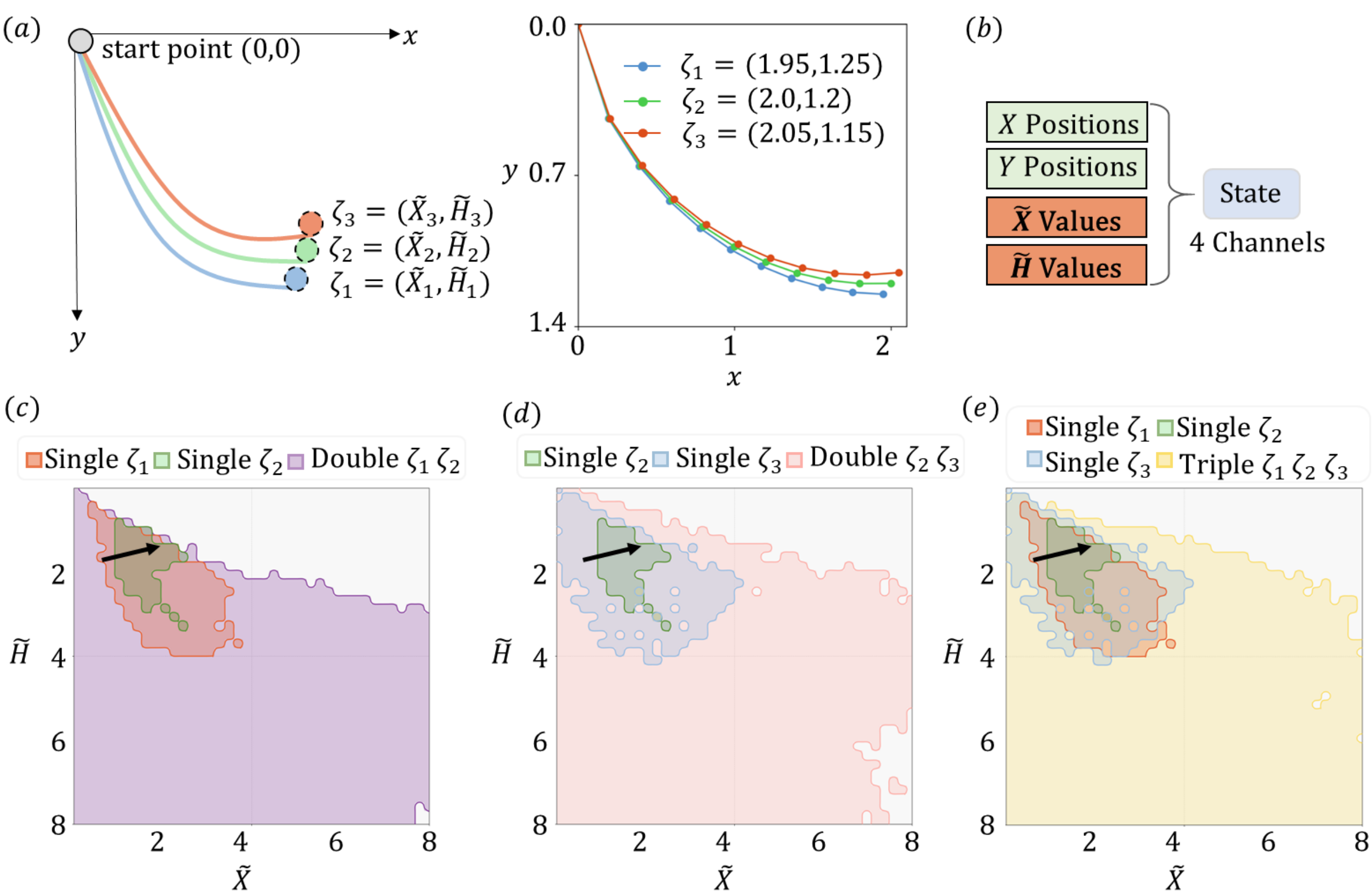

Figure 3. Generalization performance of episodic observation-switching training in the brachistochrone problem. (a) Brachistochrone problem with three closely spaced terminal points. The left panel illustrates the schematic setup with a common starting point (0,0), while the right panel shows the corresponding exact brachistochrone trajectories for the three training observations $\zeta_1$, $\zeta_2$, and $\zeta_3$, corresponding to $(\tilde{X}, \tilde{H}) = (1.95,1.25)$, $(2.00,1.20)$, and $(2.05,1.15)$, respectively, with gravity acting downward. (b) State representation for the brachistochrone problem, consisting of four channels: collocation coordinates along the $x$- and $y$-directions $(X, Y)$, and the terminal point coordinates $(\tilde{X}, \tilde{H})$. (c–e) Generalization performance comparison between single-observation training and episodic observation-switching training. The color-shaded area denotes the region with $R^2 \geq 0.90$. (c) Generalization from individual observations $\zeta_1$ and $\zeta_2$ versus switching training on $\zeta_1, \zeta_2$. (d) Analogous comparison for observations $\zeta_2$ and $\zeta_3$. (e) Switching training on three observations $\zeta_1, \zeta_2, \zeta_3$ yields high-accuracy predictions over a substantially enlarged parameter domain, indicating a generalization gain beyond that achievable by individual observations. The black arrows in (c–e) indicate the training observations.

### 3.2 Large-Deformation Nonlinear Elastic Plate

We next validate the proposed "train–freeze–execute" protocol on a large-deformation square plate problem to test the method's applicability to real-world computational mechanics tasks. Nonlinear geometric strains are considered. The physical setup is illustrated in Figure 4(a): a square plate of side length $\tilde{B}$ and orientation angle $\tilde{\Theta}$ is clamped along its boundary and subjected to a uniform transverse load $\tilde{P}$ perpendicular to the plate surface. The agent's objective is to predict the full displacement field of the plate, both in-plane components along the $x$-axis $U(x, y)$ and along the $y$-axis $V(x, y)$, as well as the out-of-plane component $W(x, y)$ under the imposed boundary condition. In our implementation the constitutive response is linear elastic (constant Young's modulus and Poisson's ratio), consistent with the Abaqus reference solutions used for high-fidelity comparison. Accordingly, the system's total potential energy $E$, defined as the internal strain energy minus the external work of the applied load, is evaluated at each step by integrating the strain energy density over the plate domain together with the load contribution. This variational principle is encoded into the RL formulation through the energy-based reward: after each action during the training rollout, the total potential energy $E_t$ of the current displacement field is computed, and a signed improvement reward is assigned as $r_t = \mathrm{sign}(E_{t-1} - E_t)$, such that actions reducing the total potential energy are positively reinforced. This energy-based reward gives the agent a physically consistent objective that aligns with the variational equilibrium principle for the stable response branch considered here.

To condition the network on task parameters, $\tilde{P}$, $\tilde{B}$ and $\tilde{\Theta}$ are placed in dedicated state channels (Figure 4(b)) so that the per-collocation-point state vector contains six channels $s = (W, U, V, \tilde{P}, \tilde{B}, \tilde{\Theta})$. Using this parameterized state representation, three distinct tasks are designed to probe the method's ability to exploit inter-task similarity. Task I (Varying Load Magnitude) concerns learning displacement fields under varying load $\tilde{P}$ with fixed plate size $\tilde{B} = 100\ \mathrm{mm}$ and orientation $\tilde{\Theta} = 0^{\circ}$, while Task II (Varying Plate Size) involves learning size-dependent responses under varying $\tilde{B}$ with fixed load $\tilde{P} = 1 \times 10^{-5}\ \mathrm{MPa}$ and orientation $\tilde{\Theta} = 0^{\circ}$, and Task III (Varying Orientation) focuses on learning orientation-dependent displacement patterns under varying $\tilde{\Theta}$ with fixed load $\tilde{P} = 1 \times 10^{-5}\ \mathrm{MPa}$ and plate size $\tilde{B} = 100\ \mathrm{mm}$. For each task, we construct a local cluster comprising three neighboring observations, denoted as $\zeta_1, \zeta_2,$ and $\zeta_3$, respectively. Figure 4(a) illustrates the degree of similarity among the three neighboring observations used in each task. Using the identical DQN backbone and hyperparameter set across variants, we train six model variants per task following the protocol described previously. Frozen models are then evaluated on dense test grids spanning large, previously unseen ranges of $\tilde{P}$, $\tilde{B}$ and $\tilde{\Theta}$ and accuracy is measured by the mean coefficient of determination $\bar{R}^2 = \frac{R_W^2+R_U^2+R_V^2}{3}$ against high-fidelity Abaqus references, with $\bar{R}^2 \geq 0.90$ denoting "generalized".

The experimental results demonstrate that the episodic switching training produces robust, observation-efficient mechanics intuition across all tasks. Across all three tasks, the agent trained with triple-observation switching accurately reproduces the reference finite-element solutions at the training configurations and, crucially, maintains high predictive fidelity when evaluated beyond the training observations within the tested parameter ranges, as illustrated by the representative predicted fields in Figures 4(c–k). Despite substantial variations in load magnitude, plate size, or orientation, the selected displacement fields remain in close agreement with the Abaqus reference solutions in terms of $\bar{R}^2$. These results demonstrate strong generalization within the tested parameter ranges for the load-, size-, and orientation-varying tasks. Figure 4(l) summarizes the $\bar{R}^2$-based generalization intervals for the elastic-plate tasks: single-observation training produces narrowly localized intervals near the training points, whereas double- and triple-observation switching yield intervals that extend beyond the union of the corresponding single-observation intervals. In all tasks, triple-observation training yields broad coverage across the sampled parameter axis, reproducing finite-element displacements with high fidelity near training points, as measured by $\bar{R}^2$, and showing predictable, gradual degradation toward the empirical parameter-interval boundaries.

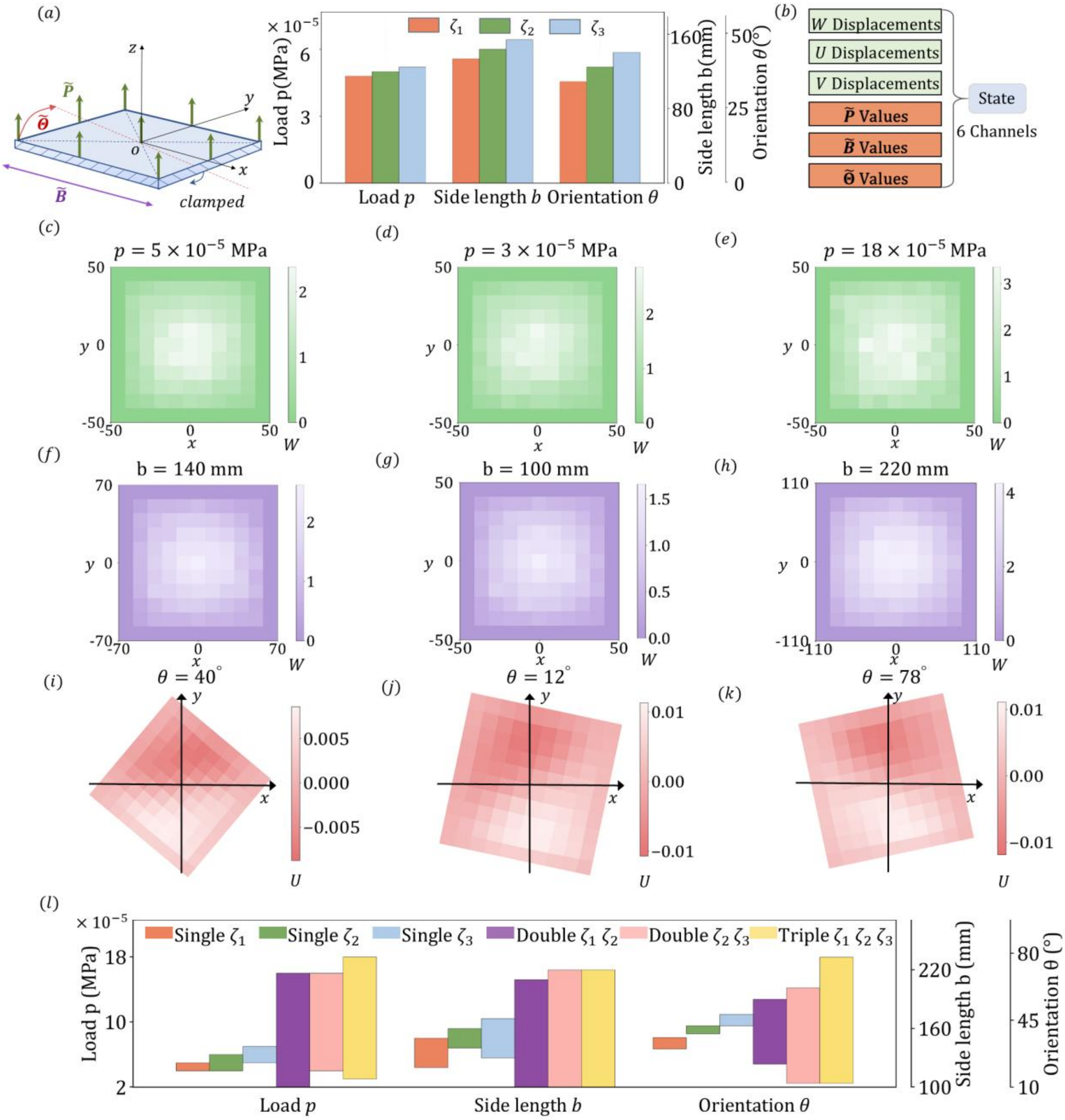

Figure 4. Generalization of episodic observation-switching learning on large-deformation elastic-plate tasks. (a) Left: schematic of the large-deformation elastic plate with clamped edges; right: three closely spaced training observations $\zeta_1$, $\zeta_2$, and $\zeta_3$ selected for the three tasks, showing only small differences between observations. (b) State representation for the elastic-plate problems, consisting of six channels: out-of-plane displacement $W$, in-plane displacements $U$ and $V$, applied load $\tilde{P}$, plate side length $\tilde{B}$, and plate orientation $\tilde{\Theta}$. (c–e) Task I (varying load): predicted $W$ distributions from a model trained on three nearby $\tilde{P}$ observations, evaluated at the training point ($\tilde{P} = 5 \times 10^{-5}$ MPa) and at two distant test points ($\tilde{P} = 3 \times 10^{-5}$ MPa and $18 \times 10^{-5}$ MPa), where the two distant predictions achieve $\bar{R}^2 \approx 0.94$ and $\bar{R}^2 \approx 0.91$, respectively. (f–h) Task II (varying plate size): predicted $W$ distributions from the model trained on three similar $\tilde{B}$ observations and evaluated at the training point ($\tilde{B} = 140$ mm) and distant evaluation points ($\tilde{B} = 100$ mm and $220$ mm), where the two distant predictions achieve $\bar{R}^2 \approx 0.94$ and $\bar{R}^2 \approx 0.92$, respectively. (i–k) Task III (varying orientation): predicted in-plane displacement $U$ from the model trained on three nearby $\tilde{\Theta}$ observations, evaluated at the training point ($\tilde{\Theta} = 40°$) and two distant orientations ($12°$ and $78°$), where both distant predictions achieve $\bar{R}^2 \approx 0.91$. (l) Quantitative summary of generalization intervals for all models in all tasks: multi-observation switching training consistently yields substantially enlarged generalization intervals relative to single-observation training, indicating an expansion of the generalization interval that exceeds the union of single-observation results.

### 3.3 Quantum harmonic oscillator

As a compact demonstration for possible application in quantum mechanics, we apply the same protocol to the quantum harmonic oscillator. The objective is to minimize the energy expectation value, i.e., the expectation value of the Hamiltonian, $\langle E \rangle = \frac{\int_{x_{min}}^{x_{max}} \psi(x)(H\psi)(x)\,dx}{\int_{x_{min}}^{x_{max}} \psi(x)^2\,dx}$, and the parameter-conditioned state is $s = (X, Y, \tilde{m}, \tilde{\omega})$, where $X$ is the spatial coordinate, $Y = \psi(X)$ is the current wavefunction value at the node, and $(\tilde{m}, \tilde{\omega})$ are the episode-specific mass and angular frequency.

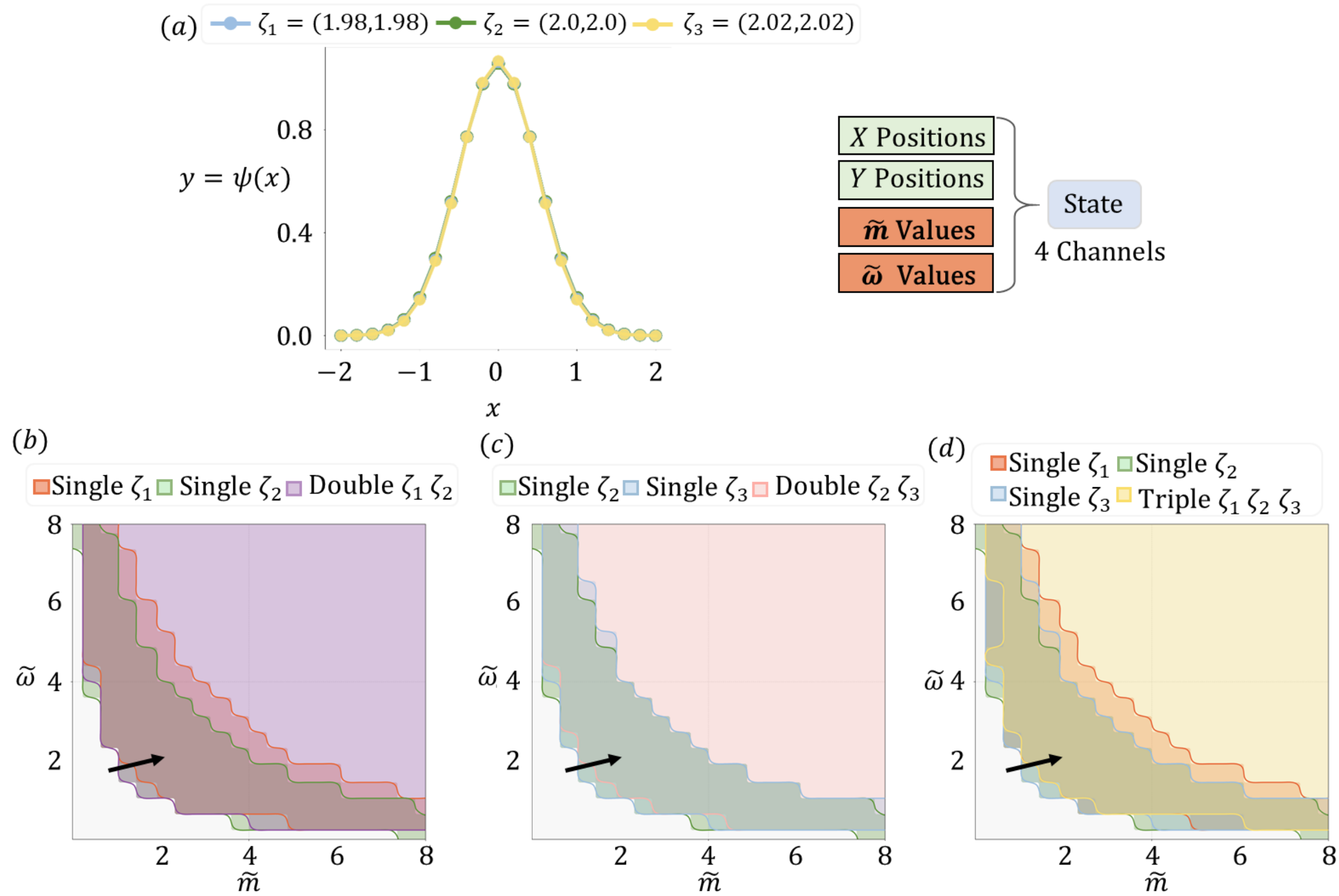

Figure 5. Generalization performance of episodic observation-switching training in the quantum harmonic oscillator problem. (a) Left: analytic ground-state wavefunctions for the three training observations $\zeta_1$, $\zeta_2$ and $\zeta_3$, corresponding to $(\tilde{m}, \tilde{\omega}) = (1.98, 1.98)$, $(2.00, 2.00)$, and $(2.02, 2.02)$, respectively; the analytical wavefunctions are shown only as reference solutions. Right: the structured state encoding used by the agent, consisting of four channels per collocation point: the spatial coordinate $X$ and the current wavefunction value $Y$, and the parameter channels $(\tilde{m}, \tilde{\omega})$. (b–d) Generalization performance comparison between single-observation training and episodic observation-switching training. The color-shaded area denotes the region with $R^2 \geq 0.90$. (b) Generalization from individual observations $\zeta_1$ and $\zeta_2$ versus switching training on $\zeta_1, \zeta_2$. (c) Analogous comparison for observations $\zeta_2$ and $\zeta_3$. (d) Switching training on three observations $\zeta_1, \zeta_2, \zeta_3$ yields an enlarged empirical generalized region on the independent test grid, defined by $R^2 \geq 0.90$, indicating a clear generalization gain relative to single-observation training. The black arrows in (b–d) indicate the training observations.

Figure 5(b–d) shows that the same trend observed in the brachistochrone and elastic-plate tasks persists: Models

trained from a single observation typically achieve $R^2 \geq 0.90$ only in the immediate neighborhood of the training point; predictive accuracy degrades rapidly as the test parameters depart from that point. By contrast, models trained with two or three similar observations produce substantially enlarged domains of $R^2 \geq 0.90$, demonstrating that episodic switching among nearby observations yields a broader generalization region. Representative comparisons confirm that, near training observations, the learned wavefunctions recover both global waveform shape and local curvature with high fidelity relative to the analytic solution, while at the edges of the empirical domain the learned solutions preserve the qualitative ground-state profile but exhibit modest quantitative deviations (reduced $R^2$ or small curvature misalignments).

These findings indicate that the episodic switching protocol can help the agent generalize across a family of related Hamiltonians. This potential connection underscores that our framework is not merely a tool for classical mechanics, but a unified approach for building parameter-conditioned intuition across the physical sciences, wherever a well-defined objective functional exists.

## 4. Unified Reinforcement Learning Theory on the Data-Efficient Generalization

To explain the observed data-efficient generalization, we provide a theoretical interpretation based on Bellman optimality, geometric learning, and algorithmic implementation. Mechanics intuition is formalized as the smooth variation of the optimal action-value function with continuous physical parameters, and our training algorithm is rationalized within the Bellman optimality framework.

### 4.1. Parameterized Markov Decision Processes

We consider a family of deterministic Markov Decision Processes $M_\zeta = (s, A, P_\zeta, r_\zeta, \gamma)$ parameterized by a physical observation parameter $\zeta$. The state $s$ is augmented as $s = (u, \zeta)$, where $u$ collects all non-parameter channels (current candidate solution and spatial descriptors). Under this representation, the agent learns a parameter-conditioned action-value function $Q_\theta(s, a)$, with $s = (u, \zeta)$, equivalently written as $Q_\theta(u, \zeta, a)$.

For each physical observation parameter $\zeta$, the corresponding Bellman optimality operator $B_\zeta$ is defined as

$$\left(B_\zeta[Q(\cdot,\zeta,\cdot)]\right)(u,a) = \mathbb{E}_{u' \sim P_\zeta(\cdot|u,a)}\left[\tilde{r}_\zeta(u,a,u') + \gamma \max_{a'} Q(u',\zeta,a')\right]. \tag{1}$$

Here, $P_\zeta$ and $\tilde{r}_\zeta(u, a, u')$ denote, respectively, the transition kernel and reward function associated with the fixed-$\zeta$ slice of the augmented environment. During a single episode, the physical observation parameter $\zeta$ is held fixed; episodic switching occurs only at episode boundaries. For the deterministic environments considered in this work, applying action $a$ to the current non-parameter state $u$ produces a unique post-action state $u' = F_\zeta(u, a)$. Hence, $P_\zeta(\cdot\,|u, a)$ is a Dirac transition kernel concentrated at $F_\zeta(u, a)$. The notation $P_\zeta$ accounts for possible parameter-dependent deterministic updates, such as the load-scaled action step sizes used in the varying-load elastic-plate problem. Thus, although the transition is deterministic, the corresponding map may depend on $\zeta$. When a normalized shared-transition approximation is used, one may set $P_\zeta \equiv P$; otherwise, the parameter dependence of $P_\zeta$ is retained through $F_\zeta$.

The numerical DQN uses a bounded sign-improvement reward, $r_\zeta^{\mathrm{num}}(u, a, u') = \mathrm{sign}[J_\zeta(u) - J_\zeta(u')]$, where $J_\zeta$ is the physical objective associated with the active observation. Because this sign map is discontinuous, the smooth perturbation analysis is not applied directly to $r_\zeta^{\mathrm{num}}$. Instead, it is stated for the continuous improvement reward $\bar{r}_\zeta(u, a, u') = J_\zeta(u) - J_\zeta(u')$, or for a bounded smooth approximation of it. Thus, $\tilde{r}_\zeta$ in Eq. (1) denotes the numerical reward in the implementation and the smooth surrogate in the theoretical analysis.

Under the standard assumptions of bounded rewards and $\gamma \in (0,1)$, each $B_\zeta$ is a contraction mapping in the sup-norm and therefore admits a unique fixed point $Q_\zeta^* = B_\zeta[Q_\zeta^*]$, which is the optimal action-value function for the specific physical observation parameter $\zeta$. This fixed-$\zeta$ statement provides the basis for comparing neighboring Bellman fixed points in the following subsection.

**4.2 Smoothness of Optimal Solutions Under Parameter Perturbations**

We compare two nearby physical observation parameters $\zeta_i$ and $\zeta_j$. Let $B_i = B_{\zeta_i}$, $B_j = B_{\zeta_j}$ denote the corresponding Bellman operators, with fixed points $Q_i^*$ and $Q_j^*$ for the associated surrogate Bellman problems. For the surrogate Bellman problem, the contraction property of the Bellman operator gives the local perturbation bound

$$\|Q_i^* - Q_j^*\|_\infty \leq \frac{C}{1-\gamma}\|\zeta_i - \zeta_j\|. \tag{2}$$

Here, $C$ is a local perturbation constant. In the normalized shared-transition case, it reduces to the Lipschitz constant of the surrogate reward with respect to $\zeta$; when the deterministic update map also depends on $\zeta$, $C$ additionally includes the locally controlled transition contribution. Eq. (2) formalizes the assumption that neighboring physical problems have neighboring optimal action-value functions. Learning a single function $Q_\theta(u, \zeta, a)$ that accurately approximates $Q_\zeta^*$ across a local range of $\zeta$ is therefore a locally meaningful learning problem, provided we can embed this smoothness prior into the learning process.

**4.3 Pitfall of Single-Task Training and the Need for Multi-Task Consistency**

Single-observation training can be idealized as reducing the Bellman residual only for the fixed parameter $\zeta_k$. With the notation of Sec. 4.1, the corresponding population-level residual loss may be written as

$$\mathcal{L}_{\text{single}}(\theta; \zeta_k) = \mathbb{E}_{(u,a)\sim\mu_k}\left[\left(Q_\theta(u, \zeta_k, a) - \left(B_{\zeta_k}[Q_\theta(\cdot, \zeta_k, \cdot)]\right)(u, a)\right)^2\right], \tag{3}$$

where $\mu_k$ denotes the state-action distribution sampled during training at $\zeta_k$. In the DQN implementation, this idealized Bellman-residual objective is approximated by sampled squared temporal-difference errors from the replay buffer.

Minimizing Eq. (3) may yield an accurate approximation $Q_\theta(\cdot, \zeta_k, \cdot) \approx Q_{\zeta_k}^*$ on the sampled region of the single task. However, it imposes no direct Bellman-consistency constraint at neighboring parameters $\zeta \neq \zeta_k$. Consequently, the learned dependence of $Q_\theta$ on $\zeta$ may be poorly controlled away from the observed task, even though the network takes $\zeta$ as an input. This is the main limitation of single-observation training: it can fit an isolated task without necessarily capturing the local solution manifold across nearby physical parameters.

We therefore view intuitive learning as an idealized multi-parameter Bellman-consistency problem: finding a shared parameter set $\theta^*$ that reduces the Bellman residuals for a set of neighboring physical parameters (e.g., $\zeta_1, \zeta_2, ...$):

$$\theta^* \in \arg\min_\theta \mathcal{L}_{\text{multi}}(\theta; \{\zeta_k\}), \qquad \mathcal{L}_{\text{multi}}(\theta; \{\zeta_k\}) = \sum_k \left\| Q_\theta(\cdot, \zeta_k, \cdot) - B_{\zeta_k}[Q_\theta(\cdot, \zeta_k, \cdot)] \right\|_\infty. \tag{4}$$

Eq. (4) is an idealized consistency objective. In the implemented DQN, episodic observation-switching approximates this requirement by alternating among neighboring $\zeta_k$ values and storing their transitions in a shared replay buffer. The sampled temporal-difference updates therefore come from an empirical mixture of nearby Bellman operators, which biases the shared network toward local multi-parameter consistency rather than specialization to one parameter value.

### 4.4 Transition to a Unified Geometric View

The preceding analysis suggests that multi-parameter Bellman consistency biases the learned parameter-conditioned action-value function toward the local regularity of the optimal action-value functions. This provides a theoretical motivation for our episodic observation-switching method. To further interpret the observed generalization capability (namely, why training on only two or three neighboring observations can yield a substantially broader region of accuracy), we examine the geometry of the underlying solution space. We view the continuum of physical tasks as a locally smooth parameterized family and interpret the learned action-value function as an approximation to a low-dimensional, manifold-like solution structure embedded in a high-dimensional function space. This provides a geometric interpretation of data-efficient, intuitive generalization.

### 4.5 The Parameter Manifold and Solution Submanifold

The physical observation parameters $\zeta$ can be regarded locally as coordinates on a parameter manifold $\mathcal{M}$. Each point $\zeta \in \mathcal{M}$ specifies a task instance, such as a brachistochrone endpoint, an elastic-plate loading condition or a harmonic-oscillator parameter pair. Under the compact state notation $s = (u, \zeta)$, the corresponding optimal action-value function $Q^*_\zeta$ is the Bellman fixed point associated with that fixed task parameter, including its parameter-dependent reward and, when present, its parameter-dependent deterministic update rule. The family of optimal action-value functions $\{ Q^*_\zeta : \zeta \in \mathcal{M} \}$ is not expected to vary arbitrarily over the full high-dimensional function space. Because the tasks are governed by related physical laws and differ only through continuous observation parameters, this family can be interpreted as lying on, or near, a lower-dimensional manifold-like solution structure. Under this view, learning mechanics intuition corresponds to approximating this latent solution structure with a shared parameter-conditioned function $Q_\theta(s, a)$, where $s = (u, \zeta)$, equivalently $Q_\theta(u, \zeta, a)$, rather than independently fitting isolated task-specific value functions.

### 4.6 Stationarity of the Bellman Residual on the Parameter Manifold

For the surrogate Bellman problem, we now state the residual-stationarity property that connects the local Bellman fixed-point structure to the finite-difference consistency encouraged by episodic observation-switching. The parameter-manifold setting, deterministic update $u' = F_\zeta(u, a)$, and possible parameter dependence of $F_\zeta$ have been introduced above. The differentiability and local non-degeneracy assumptions are needed for this residual-stationarity argument.

For the differentiable perturbation analysis only, $\bar{r}_\zeta$ denotes the smooth surrogate reward associated with the objective improvement $J_\zeta(u) - J_\zeta(u')$. It may be the raw improvement functional restricted to a compact reachable domain, provided it is uniformly bounded and locally Lipschitz, or a bounded smooth approximation of it, such as a clipped or tanh-smoothed improvement reward.

Since the environments are deterministic, $P_{\zeta(\cdot|u,a)}$ is concentrated at $u' = F_\zeta(u, a)$. Therefore, the equation reduces to

$$Q^*(u,\zeta,a) = \bar{r}_\zeta\left(u, a, F_\zeta(u,a)\right) + \gamma \max_{a'} Q^*\big(F_\zeta(u,a), \zeta, a'\big), \tag{5}$$

where $Q^*(u,\zeta,a)$ denotes the parameter-indexed optimal action-value function $Q_\zeta^*(u,a)$ and $\gamma \in (0,1)$ is the discount factor.

For any candidate $Q$, define the corresponding surrogate Bellman residual as

$$\delta_Q(\zeta; u, a) = Q(u,\zeta,a) - \bar{r}_\zeta\left(u, a, F_\zeta(u,a)\right) - \gamma \max_{a'} Q\big(F_\zeta(u,a), \zeta, a'\big). \tag{6}$$

The exact fixed-point family therefore satisfies

$$\delta_{Q^*}(\zeta; u, a) = 0, \tag{7}$$

for every admissible $(u, a)$ and each fixed $\zeta$. To obtain the residual-stationarity identity, perturb $\zeta$ along a tangent direction $v \in T_\zeta\mathcal{M}$. Under local differentiability of $Q^*$, $\bar{r}_\zeta$, and $F_\zeta$, and away from action-tie or policy-switching regimes so that the locally maximizing action remains stable, differentiating Eq. (7) gives

$$D_v \delta_{Q^*}(\zeta; u, a) = 0. \tag{8}$$

Here, $D_v$ denotes the total directional derivative induced by the parameter perturbation, including the induced variation of the deterministic post-action state $F_\zeta(u, a)$ when $F_\zeta$ depends on $\zeta$. Thus, for the exact surrogate Bellman fixed-point family, the residual is identically zero and therefore stationary along the parameter manifold. This is a reference identity for the ideal fixed-point family, not a proof that the implemented DQN explicitly minimizes residual derivatives across parameters. This identity should be viewed as a consistency reference for the learned approximation, rather than as a direct consequence of the training algorithm. This stationarity is not an additional assumption; it follows from differentiating the Bellman optimality equation along a local parameter perturbation. The full Taylor expansion, the local action-gap condition for the active maximizer, and the chain-rule terms associated with parameter-dependent deterministic updates, are utilized for derivation.

During learning, however, the network represents an approximation $Q_\theta$, whose residual is generally nonzero. A learned approximation that tracks the neighboring fixed-point family should not only keep the residual small at each sampled parameter, but also avoid sharp residual variations across nearby parameters. This motivates the heuristic consistency condition

$$D_v \delta_{Q_\theta}(\zeta; u, a) \approx 0. \tag{9}$$

This condition should not be interpreted as an exact constraint imposed by the learning algorithm. Rather, it describes the local residual regularity expected from a parameter-conditioned approximation that follows the neighboring optimal solutions.

Episodic observation-switching provides a discrete mechanism for encouraging this regularity. Because different components of the physical parameter $\zeta$ may have different units and numerical scales, all finite differences below are evaluated in the normalized parameter coordinate $\hat{\zeta}$, obtained by linearly rescaling each parameter component to a common interval, such as [0,1]. This normalization makes the denominator dimensionally consistent and prevents the quotient from depending on arbitrary choices of physical units. For two nearby parameters $\zeta_i$ and $\zeta_j$, the vanishing directional variation of the Bellman residual along the chord connecting them can be estimated by

$$\frac{\left\| \delta_{Q_\theta}(\hat{\zeta}_i; \cdot, \cdot) - \delta_{Q_\theta}(\hat{\zeta}_j; \cdot, \cdot) \right\|_\infty}{\left\| \hat{\zeta}_i - \hat{\zeta}_j \right\|} \approx 0 \tag{10}$$

Here $\hat{\zeta}$ denotes the normalized parameter coordinate used by the network. The use of $\hat{\zeta}$ avoids dependence of the quotient on arbitrary physical units or parameter scaling. The residuals in Eq. (10) are compared after all task-dependent states have been embedded into the same normalized computational representation; otherwise the sup-norm difference is not well defined. Episodic switching does not directly minimize this finite-difference quotient. Rather, by reducing sampled temporal-difference errors at neighboring parameters with a shared parameter-conditioned network, it provides a discrete mechanism that can indirectly discourage sharp residual variations. The implementation-level mechanism is discussed in Sec. 4.7. Thus, Eq. (10) serves as a finite-difference expression of the local Bellman-consistency principle that underlies the generalization mechanism discussed below.

### 4.7 Algorithmic Mechanism: Experience Mixing as Implicit Regularization

The episodic switching protocol has a critical implementation-level consequence: the shared experience replay buffer $\mathcal{D}$ is populated with transition tuples $(s, a, r, s')$ collected under different physical observation parameters $\zeta_i$ and $\zeta_j$. Since $\zeta$ is encoded in the state channels, each stored transition carries the task context under which it was generated. Consequently, when a minibatch is sampled from $\mathcal{D}$ for Q-network optimization, it naturally constitutes a mixture of experiences from neighboring tasks. In expectation, the resulting temporal-difference updates are directed toward reducing sampled Bellman-consistency errors over multiple neighboring $\zeta$ values, rather than over a single isolated task. This experience mixing biases the optimization against network-parameter settings that fit one observation well but give inconsistent value estimates at nearby observations. Instead, the shared network $Q_\theta$ is encouraged to represent a unified parameter-conditioned action-value function that approximately satisfies the Bellman optimality conditions over the local task cluster. Thus, episodic observation-switching acts as an implicit multi-task regularization mechanism. It provides an algorithmic route for encouraging the finite-difference residual-consistency behavior described around Eq. (10) and supports the geometric view of learning a coherent function over the parameter manifold.

### 4.8 Explaining the Generalization Gain

This unified view provides a theoretical interpretation of the key computational findings. Single-observation training generalizes poorly because it reduces the Bellman residual, or, in implementation, the sampled temporal-difference residual, only at an isolated point on $\mathcal{M}$, with no direct Bellman-consistency constraint at neighboring parameters. The learned parameter-conditioned action-value function $Q_\theta$ is therefore weakly constrained by the local solution structure, leading to rapid degradation away from the training observation.

Switching among a few similar observations improves generalization because training on a local cluster $\zeta_1, \zeta_2, \zeta_3$ encourages finite-difference residual consistency within that neighborhood, as described by Eq. (10). This biases $Q_\theta$ toward the local tangent structure of the solution submanifold. When the local geometry is sufficiently captured by the shared approximator, the learned function can generalize over a broader contiguous region of the parameter manifold, provided the tested parameters $\zeta_{test}$ remain within the same regular physical regime. Under these empirical and regularity conditions, the generalization region can expand beyond the union of isolated point neighborhoods, as observed in the tested cases, covering a continuous portion of $\mathcal{M}$. This mechanism provides a theoretical interpretation of the substantial expansion of the high-accuracy regions observed in our computational experiments: training on a local cluster of neighboring tasks yields generalized regions or intervals that extend beyond the corresponding single-observation results in the tested cases.

### 4.9 Connection to Broader Themes

Notably, the generality of our framework extends well beyond classical mechanics (e.g., the quantum oscillator problem as discussed earlier). The core methodology, encoding continuous physical parameters into the state and using the physical functional as the reward signal, can in principle be applied to optimization problems governed by computable objective functionals, provided that suitable state representations and action spaces can be designed. This includes diverse fields such as electromagnetics (e.g., field-shaping for minimal radiation loss), thermodynamics (e.g., heat conduction path optimization), and fluid dynamics (e.g., drag minimization in shape design).

## 5 Conclusions

By encoding physical parameters into the state and switching between related tasks during training, our method enables the agent to acquire broad mechanics intuition from a few examples, as demonstrated on two classical mechanics problems and a quantum mechanics example. Theoretically, this is associated with encouraging approximate stationarity of the Bellman residual across parameters, which biases the learner toward a consistent value function over the task family.

Geometrically, our method learns a smooth function over the task parameter manifold, supporting generalization within the tested regular regimes. This success stems from two synergistic components: a physics-informed smoothness prior, instantiated by sampling closely related tasks, and the episodic switching protocol, which mixes experiences across tasks and acts as an implicit regularizer. The prior defines the low-dimensional solution structure to be learned; the switching protocol enables its acquisition from scarce data. Ultimately, the learning process is biased so that small task variations tend to preserve Bellman consistency, naturally promoting generalization.

These findings offer a computational perspective on how data-efficient mechanics intuition might emerge in both artificial and biological systems. In cognitive science, human generalization is often described as the abstraction of conceptual models from sparse examples [5]. While the underlying mechanisms are undoubtedly different, our RL agent exhibits a functionally similar capability: by encoding parameters and encouraging Bellman consistency, it develops a parameter-conditioned policy that empirically generalizes across the tested regular portions of the task family, as if it had formed an abstract, parameterized model of the underlying physics. The resulting generalization should be interpreted within local regular regimes where the objective functional, transition map, and optimal policy vary smoothly with the task parameter. This suggests that encouraging consistency across related experiences, whether through cognitive abstraction or through algorithmic biases toward approximately stationary Bellman residuals, may be a general principle for efficient learning.

Thus, our work not only demonstrates strong generalization in physical task families but also provides a unifying computational principle: maintaining coherence across related tasks is key to extracting broad insights from limited data. This principle aligns with insights from neuroscience-inspired AI, where structural biases (e.g., recurrent circuits [22] ) can also promote generalization. Future research could explore whether such algorithmic principles can inform our understanding of biological learning, or conversely, how biological insights might further refine data-efficient RL. While our approach requires an objective physical functional, its integration with data-driven physics discovery [23,24] (e.g., PINNs [25–29], symbolic regression [30,31]) may help address scenarios in which the functional of interest must first be discovered or approximated from data [32,33] and the associated parameter-conditioned policy must then be learned. This hybrid paradigm could extend our method to problems in which the governing functional must first be inferred or approximated from data, thereby bridging data-driven physics discovery and model-based RL control.